\documentclass[twocolumn,prl,showpacs]{revtex4}

\usepackage{graphicx,amsmath,amssymb,hyperref}

\begin{document}

\title{Saturated-absorption cavity ring-down spectroscopy}

\author{G.~Giusfredi}
\author{S.~Bartalini}
\author{S.~Borri}
\author{P.~Cancio}
\author{I.~Galli}
\author{D.~Mazzotti}\email[e-mail: ]{davide.mazzotti@ino.it}
\author{P.~De~Natale}
\affiliation{Istituto Nazionale di Ottica (INO) - CNR, Largo Fermi 6, 50125 Firenze FI, Italy\\
and European Laboratory for Nonlinear Spectroscopy (LENS), Via Carrara 1, 50019 Sesto Fiorentino FI, Italy}

\date{\today}

\begin{abstract}
We report on a novel approach to cavity-ring-down spectroscopy with the sample gas in saturated-absorption regime. This technique allows to decouple and simultaneously retrieve empty-cavity background and absorption signal, by means of a theoretical model that we developed and tested. The high sensitivity and frequency precision for spectroscopic applications are exploited to measure, for the first time, the hyperfine structure of an excited vibrational state of $^{17}$O$^{12}$C$^{16}$O in natural abundance with an accuracy of a few parts in $10^{-11}$.
\end{abstract}


\maketitle


The availability of a molecular-spectroscopy technique, able to combine the ultimate performance in terms of sensitivity, resolution and frequency accuracy, can be crucial in many fundamental physical measurements. Indeed, profiting from the strength and ease of saturation of many mid-IR ro-vibrational transitions, this technique could provide new insights in elusive quantum-mechanical effects encoded in molecules, such as: parity violation due to weak interactions\cite{letokhov2,daussy2,faglioni2}, violation of the Fermi/Bose statistics and/or the symmetrization postulate of quantum mechanics\cite{greenberg5,modugno7,mazzotti4}, time variation of fundamental physical constants, e.g. the proton-to-electron mass ratio\cite{schiller2,reinhold}. Such a technique could also represent a major step forward in trace-gas sensing.

Concerning sensitivity, cavity ring-down (CRD) spectroscopy\cite{okeefe,romanini2} has proven to be a good candidate technique to directly provide a quantitative measurement of gas absorption coefficient with a simple experimental set-up. In principle, it is not limited by amplitude noise of the laser source, but only by detection shot noise. However, variations of the empty-cavity decay rate always prevent to achieve this ultimate limit and to average measurements over long times. The empty-cavity background could be subtracted by quickly switching the radiation frequency between nearby longitudinal cavity modes. Nevertheless, this method cannot be effective, since we observed that decay times of different cavity modes are affected by uncorrelated fluctuations. Other techniques (e.g. CRD heterodyne spectroscopy\cite{ye3} and NICE-OHMS\cite{ye}) are more sensitive than standard CRD, but they all are more complex and require fast and sensitive detectors, generally unavailable in the mid IR.

In this work we propose a new spectroscopic technique, namely saturated-absorption cavity ring-down (SCaR), that improves the CRD sensitivity. We show that the decrease of the saturation level during each SCaR event makes our technique very effective in identifying and decoupling any variation of the empty-cavity decay rate. Saturation effects have already been observed in CRD experiments and, in some conditions, they induce a deviation of the ring-down signal from the perfectly exponential behaviour. Whenever non-linear absorption occurred in CRD spectroscopy, either it was considered as a disadvantage to be avoided\cite{lehr,labazan}, or it was only used to get Lamb dips as frequency markers for line centers, without a detailed treatment of saturation effects\cite{lisak}. In some continuous-wave CRD experiments non-linear effects were observed and discussed in two specific regimes. When the ring-down time is much shorter than the population relaxation time, absorption loss keeps constant at the saturated value existing at the CRD start time and the decay will be simply exponential\cite{bucher}. In the opposite regime (the so-called ``adiabatic'' approximation), the saturation level of the optical transitions reaches the steady state at each point of the ring-down profile and a more complete and complex theoretical model would be needed to fit the non-exponential decay curves\cite{romanini7}. Theoretical mean-field analyses on dynamic absorption saturation in pulsed CRD were performed a few years ago, either considering inhomogeneous broadening\cite{lee4} or not\cite{brown2}.


We developed and tested a new model which is very effective in exploiting the SCaR spectroscopic technique. We are assuming that the gas interacts with intra-cavity radiation in a TEM$_{00}$ mode with a time-dependent intensity $I$ and power $P$ given by the following expressions:
\begin{equation}
	I(\rho,t)=I_0(t)e^{-2\left(\frac{\rho}{w}\right)^2}, \quad P(t)=\frac{\pi w^2}{2}I_0(t)
	\label{eq:i}
\end{equation}
where $\rho\equiv\sqrt{x^2+y^2}$ is the radial coordinate, $I_0(t)\equiv I(\rho=0,t)$ is the peak intensity on the cavity axis $z$ and $w$ is the beam waist, assumed to be constant along $z$. We also define several physical quantities:
\begin{subequations}
	\begin{align}
		I_s &= \frac{ch}{3}\frac{\Gamma^2 k^3}{A} &&\text{sat. intensity} \\
		P_s &= \frac{\pi w^2}{2}I_s &&\text{sat. power} \\
		G(t) &\equiv \frac{I_0(t)}{I_s}=\frac{P(t)}{P_s} &&\text{sat. parameter}
	\end{align}
	\label{eq:def}
\end{subequations}
where $A$ is the Einstein coefficient of the transition, $k$ is the wave vector and $\Gamma$ is the full width at half maximum (FWHM) of the transition. We are neglecting the effects of the standing-wave light field inside the cavity and we are spatially averaging the different saturation levels of molecules interacting with light in node and anti-node positions. In the presence of inhomogeneous broadening due to a thermal Gaussian distribution of molecular velocities, the absorption coefficient is affected by saturation and obeys the following equation\cite{demtroder}:
\begin{equation}
	\alpha(\rho,t)=\frac{\alpha_0}{\sqrt{1+I(\rho,t)/I_s}}
	\label{eq:alpha}
\end{equation}
where $\alpha_0$ is the non-saturated value of $\alpha$. Power attenuation due to gas absorption, along the $z$ axis, can be expressed as:
\begin{equation}
	\begin{split}
		\frac{dP}{dz}(t) &= -2\pi\int_0^{\infty}\alpha(\rho,t)I(\rho,t)\rho d\rho \\
		&= -\alpha_0\frac{2P(t)}{1+\sqrt{1+P(t)/P_s}}
	\end{split}
	\label{eq:dpdz}
\end{equation}
where the spatial integration over the transverse beam profile corresponds to a ``local'' approximation. That is correct at pressure regimes where the diffusion time of molecules is longer than the population relaxation time of molecular levels. Let us combine this continuous loss mechanism (following the Beer--Lambert law) with the discrete mirror losses (2 each cavity round trip), and define the decay rates $\gamma_c=c(1-R)/l$ and $\gamma_g=c\alpha_0$, where $R$ is the mirror reflectivity. Then, by using the speed-of-light relation $\frac{d}{dz}=\frac{1}{c}\frac{d}{dt}$, we get the following rate equation for the saturation parameter:
\begin{equation}
	\frac{dG}{dt}(t)=-\gamma_c G(t)-\gamma_g\frac{2G(t)}{1+\sqrt{1+G(t)}}, \quad G(0)=G_0
	\label{eq:dsdt}
\end{equation}
where $G_0=P_0/P_s$ is the saturation parameter value when the SCaR event starts, triggered by a threshold set on the signal detected in transmission. An analytical solution of this equation can be found in the form $t=t(G)$ that, unfortunately, cannot be inverted explicitly in the form $G=G(t)$. Instead, we chose to perform a numerical integration of Eqs.~\eqref{eq:dsdt} with the $4^{th}$-order Runge-Kutta algorithm (RK4) within the fitting procedure itself. Since $\gamma_g<\gamma_c$ and the dynamic range of the decay curve exceeds 4 decades, in order to get a better computing precision let us factorize $G$ as follows:
\begin{equation}
	G(t)\equiv G_0e^{-\gamma_c t}f(t;G_0,\gamma_c,\gamma_g)
	\label{eq:s}
\end{equation}
With a direct substitution in Eq.~\eqref{eq:dsdt}, we can easily find that the function $f$ obeys the following differential equation:
\begin{equation}
	\frac{df}{dt}(t)=-\gamma_g\frac{2f(t)}{1+\sqrt{1+G_0e^{-\gamma_c t}f(t)}}, \quad f(0)=1
	\label{eq:dfdt}
\end{equation}


The experimental set-up\cite{galli}, is based on a difference-frequency-generated CW coherent source widely tunable in the mid IR, with the near-IR pump/signal lasers phase-locked one another through a fs Ti:sapphire optical frequency comb (OFC)\cite{maddaloni6}. The 1-m-long cavity is formed by 2 high-reflectivity mirrors with 6-m radius of curvature and optical losses of 440~ppm around 2340~cm$^{-1}$. With this set-up we performed several spectroscopic measurements to test both sensitivity and resolution using the newly developed model. Each spectrum is recorded by stepwise scanning the absolute frequency of the near-IR signal (and thus of the mid-IR idler). At each IR frequency the cavity length is dithered at about 1~kHz rate across the resonance condition. When a fixed threshold level is reached during each cavity build-up, the IR frequency is rapidly switched off-resonance. Several CRD events are detected by a liquid-N$_2$-cooled InSb photo-diode aligned with the cavity-transmitted light, acquired by a 18-bit digitizing oscilloscope with a sampling rate of 10~MS/s, averaged by a LabVIEW acquisition program and processed on-line by a FORTRAN fitting routine.

\begin{figure}[htbp]
	\begin{center}
		\includegraphics[width=\columnwidth]{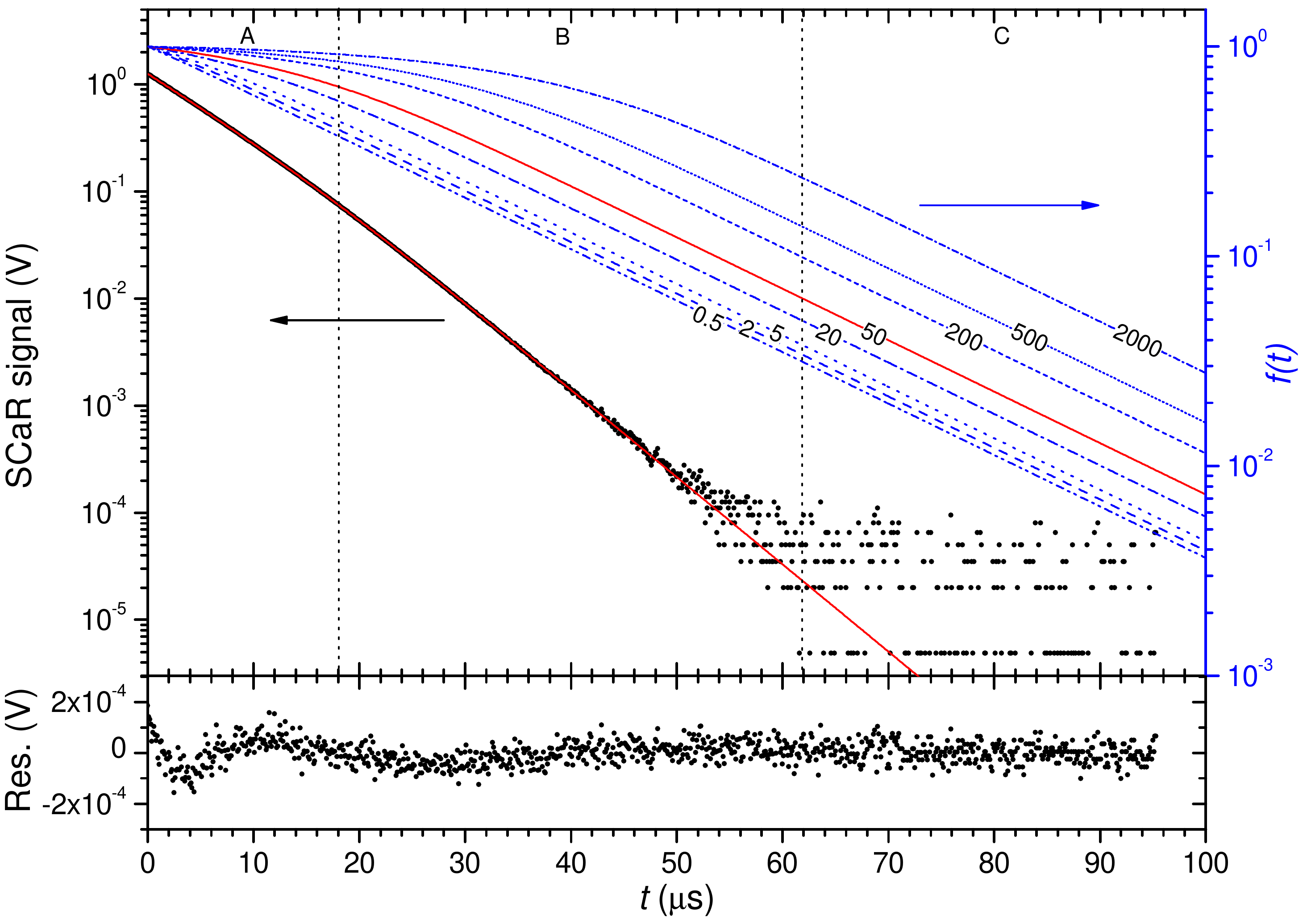}
	\end{center}
	\caption{Comparison between experimental data and theoretical model. Experimental data points for the case $G_0=50$, the fit curve (in red) and the residuals are plotted with left scale (in black). Saturated decay functions $f$, simulated for different values of $G_0$ (labeling curves), are plotted with right scale (in blue).}
	\label{fig:sat_decay}
\end{figure}

\begin{figure}[htbp]
	\begin{center}
		\includegraphics[width=\columnwidth]{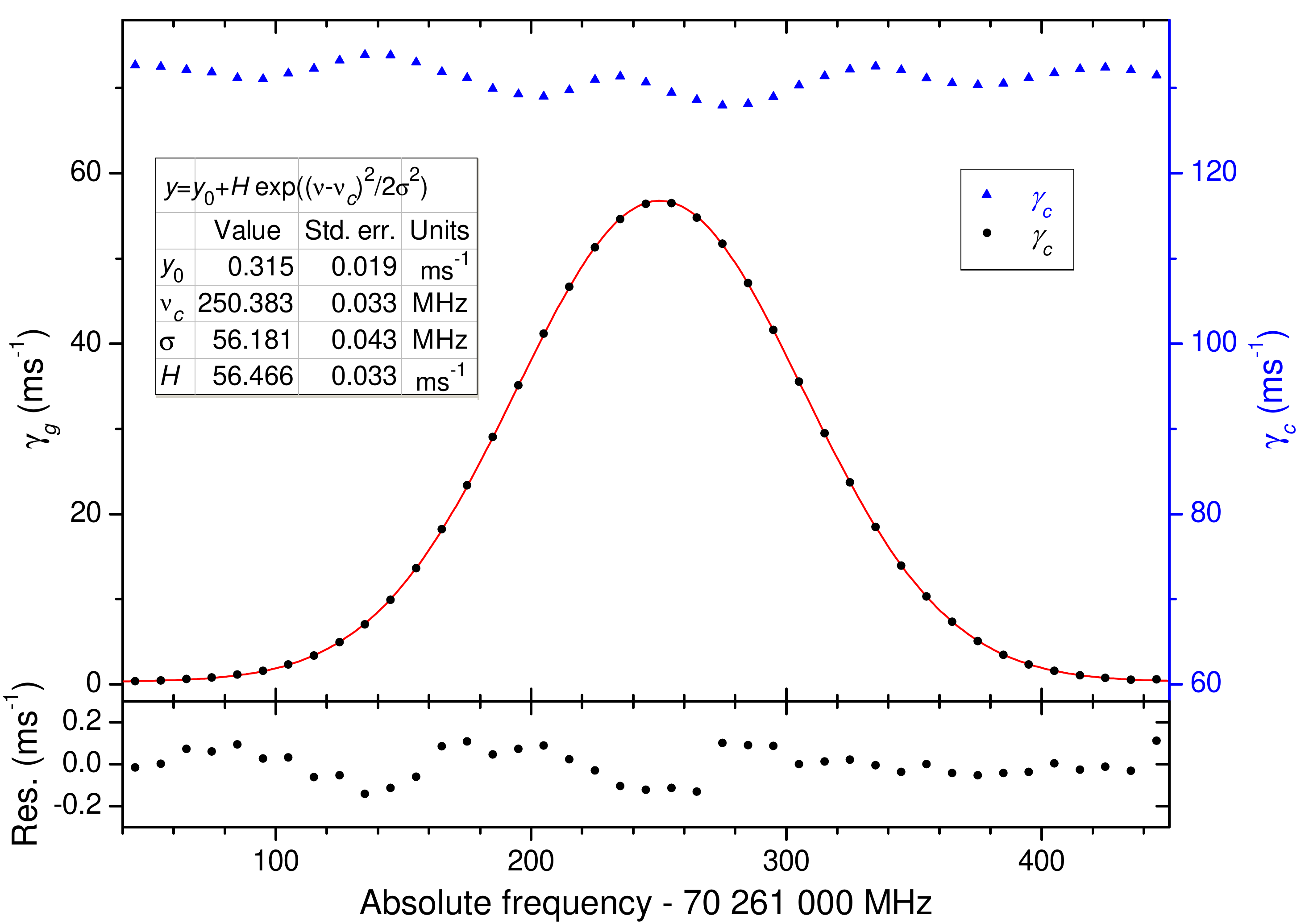}
	\end{center}
	\caption{Doppler-limited absorption spectrum of the ($03^31-03^30$) R(50) transition of $^{12}$C$^{16}$O$_2$ (line-center frequency $\nu=2343.663$~cm$^{-1}$, line-strength $S=7.86\cdot10^{-24}$~cm), recorded with absorption length $L=1$~m, temperature $T\simeq296$~K and pressure $p=50\:(2)$~$\mu$bar. The decay rate $\gamma_g$ ($\gamma_c$) is plotted with black (blue) scale at the left (right). The red curve fits $\gamma_g$ experimental data points to a Gaussian (parameters in the inset and residuals at the bottom).}
	\label{fig:Doppler}
\end{figure}

In Fig.~\ref{fig:sat_decay} SCaR measurements are compared with the model we have developed.
The experimental data in this figure are the average of 3072 decay signals measured at a fixed frequency near the absorption peak of the transition of Fig.~\ref{fig:Doppler} over a 4-s time interval. The discrete noise values in the tail of the SCaR signal are due to the resolution of the digitizing oscilloscope. Almost flat residuals witness the  validity of the theoretical model. A rough and intuitive explanation for the intrinsic ability of this technique to distinguish between empty-cavity and gas-induced decay rates can be given as follows. Three consecutive intervals can be recognized in the SCaR signal: zones A, B and C carry information respectively on $\gamma_c$, $\gamma_c+\gamma_g$ and the detector offset due to thermal background (whose fitted value has been subtracted from the signal in Fig.~\ref{fig:sat_decay}). The A-B transition is marked by a slope change in the SCaR signal, which needs the condition $G_0\gg1$ to be well observable, as it is more evident in the $f$ curves. As a consequence, a large detection dynamics is needed to measure the transition from strong-saturation to linear-absorption regime in the SCaR decay, before it falls below the noise level (B-C transition). The best choice for $G_0$ should give such three zones with similar durations.

Fig.~\ref{fig:Doppler} shows a low-pressure, Doppler-limited absorption spectrum of a ro-vibrational transition of $^{12}$C$^{16}$O$_2$ recorded with SCaR.
The free parameters in the fitting routine were $P_0,\gamma_c,\gamma_g$, while $P_s$ was kept fixed, because it is frequency independent. The non-zero background absorption is due to Lorentzian wings of nearby transitions. This spectrum was recorded in about 12~minutes by separately averaging the $\gamma_g$ and $\gamma_c$ values resulting from the SCaR fitting routine. Since the IR frequency was scanned forward and backward in 10-MHz steps, no sub-Doppler features are visible at line center.  The measured peak absorption rate $H=56.466\:(33)$~ms$^{-1}$ is consistent with the value $63$~ms$^{-1}$ calculated from the line-strength $S$ reported by the HITRAN database\cite{rothman7,hitran}, within the 5--10\% uncertainty on $S$ and the 4\% uncertainty on measured pressure. A minimum detectable absorption coefficient $\alpha_{min}=1.1\cdot10^{-9}$~cm$^{-1}$ can be derived dividing by $c$ the uncertainty value of $H$ yielded by the fitting routine. When comparing Gaussian fit residuals of $\gamma_g$ in Fig.~\ref{fig:Doppler} with the measured $\gamma_c$, it is evident that our technique effectively makes the measured gas absorption almost unaffected by any variations of the empty-cavity decay rate. Indeed, the fluctuations of the former curve are about a factor of 20 lower than the latter one and only a slight residual correlation is observed. This factor also quantifies the sensitivity improvement over standard CRD, that can only measure an overall decay rate $\gamma=\gamma_g+\gamma_c$ and is inevitably limited by any background fluctuation contained in $\gamma_c$. As mentioned above, this key point would enable a further gain of sensitivity with measurements averaged over longer times.

\begin{figure}[htbp]
	\begin{center}
		\includegraphics[width=\columnwidth]{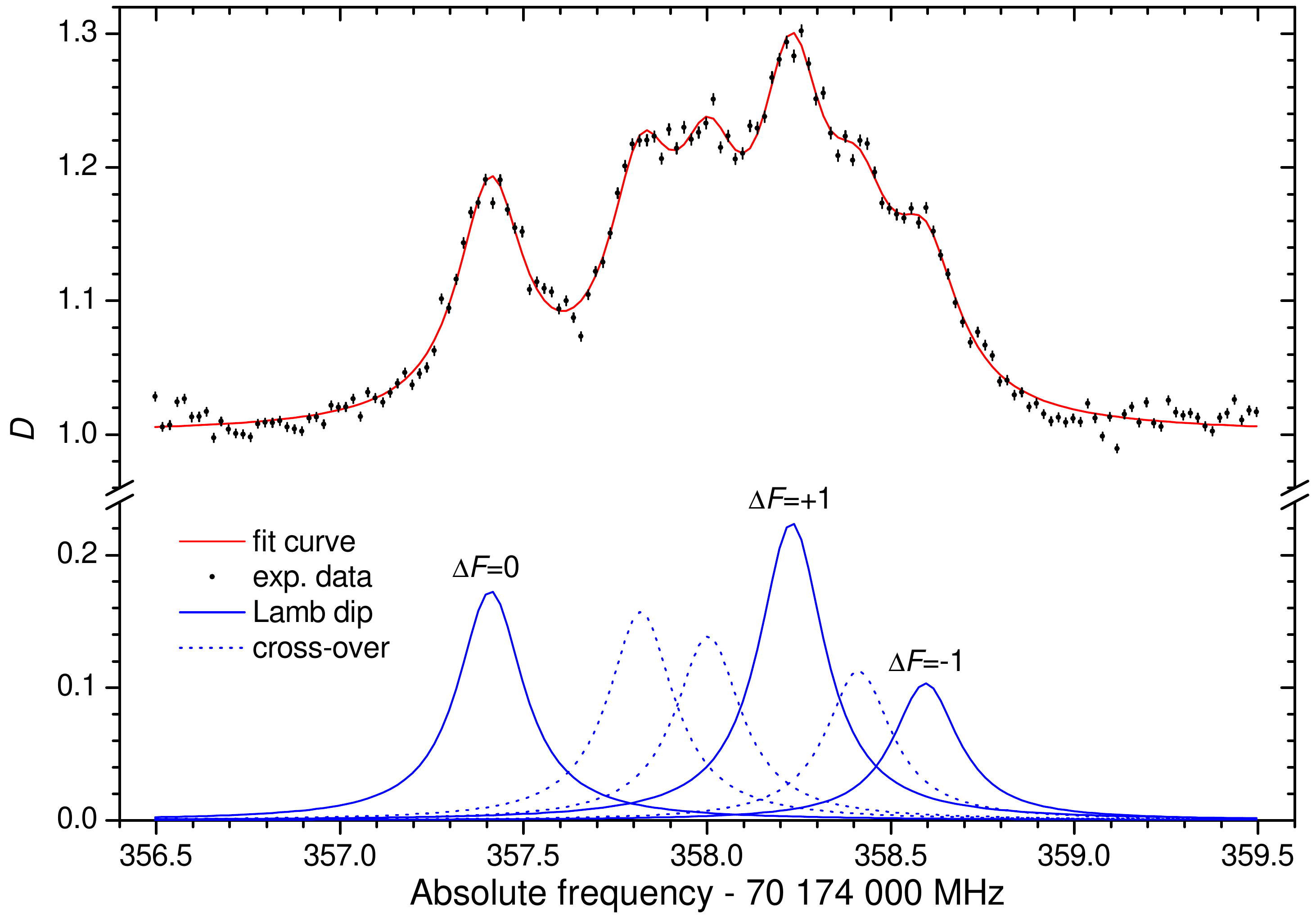}
	\end{center}
	\caption{Sub-Doppler spectrum of the ($00^01-00^00$) R(0) transition of $^{17}$O$^{12}$C$^{16}$O ($\nu=2340.765$~cm$^{-1}$, $S=1.25\cdot10^{-22}$~cm), recorded with $L=1$~m, $T\simeq296$~K and $p\sim2$~$\mu$bar. The dimensionless quantity $D$ displaying the Lamb-dip features is defined as $D\equiv G_0P_s/P_0$. The experimental data and a multi-Lorentzian curve fitting the 3 Lamb dips and the corresponding 3 cross-overs, are plotted.}
	\label{fig:hyp_struct}
\end{figure}

To test the resolution, we performed sub-Doppler measurements on low-pressure $^{17}$O$^{12}$C$^{16}$O at natural abundance ($7.5\cdot10^{-4}$). Fig.~\ref{fig:hyp_struct} shows the resolved hyperfine structure of the ($00^01-00^00$) R(0) transition of $^{17}$O$^{12}$C$^{16}$O, due to interaction between the $^{17}$O electric quadrupole (which is non-null for a nuclear spin $I=5/2$) and the electric field gradient at the nucleus position\cite{townes}.
This spectrum was recorded in about 3~hours with 11 forward/backward frequency scans in 20-kHz steps. The experimental conditions, in this case, did not satisfy the local assumption of our model, since at such a low pressure the mean free path is larger than the beam diameter and the population relaxation time is dominated by the transit time $\tau_t$. Therefore we dropped the spatial integration over the transverse beam profile of Eq.~\ref{eq:dpdz} and made a mean-field approximation. Moreover, the adiabatic approximation $\tau_t\ll1/\gamma_c$ is at the validity edge. Nevertheless, this simplified model resolved well the hyperfine structure. In this case the free parameters were $P_0,\gamma_c,D/P_s\equiv G_0/P_0$ while $\gamma_g$ was kept fixed, because the gas absorption is almost constant at the Doppler peak value during the narrow frequency scan. We report in Table~\ref{tab:hyp_struct} the fit results for the line centers.
\begin{table}[htbp]
	\begin{center}
		\begin{tabular}{c|c|c}
			$\Delta F$ & $\nu_{\Delta F}$ (kHz) & rel. intens. \\
			\hline
			-1 & $70\,174\,358\,594.9\:(6.2)$ & $0.207\:(11)$ \\
			0 & $70\,174\,357\,409.8\:(3.0)$ & $0.345\:(9)$ \\
			+1 & $70\,174\,358\,229.8\:(3.6)$ & $0.448\:(13)$
		\end{tabular}
	\end{center}
	\caption{Measured absolute frequencies $\nu_{\Delta F}$ of the hyperfine triplet with 1-$\sigma$ uncertainties and relative intensities.}
	\label{tab:hyp_struct}
\end{table}
Systematic errors due to the $10^{-12}$ relative accuracy of the absolute frequency scale (70~Hz) and to pressure-shift effects (-200~Hz) are negligible in the error budget, when compared with the statistical uncertainty (a few kHz) of the fitting routine. The relative intensities of the triplet components are nearly consistent with the theoretical 2:3:4 ratios yielded by the $I,J,F$ quantum numbers involved\cite{townes}. The fitted FWHM of each Lorentzian dip is $\Gamma=217.1\:(5.5)$~kHz. The contributions to FWHM in our experimental conditions ($w=1.5$~mm, $p\approx2$~$\mu$bar) are natural (22~Hz), pressure (14~kHz), and interaction-time broadening. The latter one is due both to transit time of molecules through the beam (55~kHz) and to mean lifetime of photons during a CRD event (21~kHz). We believe that the discrepancy with the larger measured value of $\Gamma$ can be ascribed to power broadening effects, due to the high average value of $G$ during the SCaR process. The line-center frequency $\nu_c$ and $eqQ$ parameters are obtained by fitting the hyperfine triplet to the theoretical expression depending on the $I,J,F$ and $I',J',F'$ quantum numbers. The measured line-center frequency $\nu_c=70\,174\,358\,037.3\:(3.9)$~kHz is consistent with the value $70\,174\,368$~MHz reported by HITRAN, within the declared 3--30~MHz uncertainty. Thus, we improved the frequency accuracy by more than 3 orders of magnitude. The measured electric-quadrupole coupling constant in the ro-vibrational state $|00^01;5/2,1,F\rangle$ is $eqQ=-3.927\:(30)$~MHz. A much earlier microwave measurement\cite{gripp} of $eqQ$ in the ground vibrational state was performed on a $^{17}$O$^{12}$C$^{16}$O isotopically enriched sample, with an uncertainty about a factor of 3 worse than our measurement in a excited vibrational state.


In conclusion, we have shown that SCaR spectroscopy is an effective novel approach to CRD. In particular, the saturation regime has proven to be generally useful for background reduction, whereas it also provides detection of sub-Doppler spectroscopic features, when higher resolution is needed. Finally, we have shown that frequency-comb-assisted SCaR spectroscopy can provide absolute frequency measurements of molecular transitions.

\begin{acknowledgments}
We wish to thank G.~Di~Lonardo and F.~Tamassia (University of Bologna, Italy) for useful discussions, R.~Ballerini (LENS, Italy) for making the cavity. This work was partially funded by Ente Cassa di Risparmio di Firenze and by CNR and the other European research funding Agencies participating to the European Science Foundation EUROCORES Program EUROQUAM-CIGMA .
\end{acknowledgments}

\bibliography{bib_archive}

\end{document}